\documentclass[%
 reprint,
 amsmath,amssymb,
 aps,
]{revtex4-2}

\usepackage{graphicx}% Include figure files
\usepackage{dcolumn}% Align table columns on decimal point
\usepackage{bm}% bold math
\usepackage{hyperref}% add hypertext capabilities
\usepackage[mathlines]{lineno}% Enable numbering of text and display math
\usepackage{romannum}

\usepackage[dvipsnames]{xcolor}
\usepackage{soul}
\usepackage{ulem}

\begin{document}

\preprint{APS/123-QED}

\title{Optical Switching of $\chi^{(2)}$ in Diamond Photonics}

\author{Sigurd Fl{\aa}gan}
\email{sigurd.flagan@ucalgary.ca}
 \affiliation{Institute for Quantum Science and Technology, University of Calgary, Calgary, AB, T2N 1N4, Canada}
\author{Joe Itoi}
\affiliation{Institute for Quantum Science and Technology, University of Calgary, Calgary, AB, T2N 1N4, Canada}
\author{Prasoon K.\ Shandilya}
\affiliation{Institute for Quantum Science and Technology, University of Calgary, Calgary, AB, T2N 1N4, Canada}
\author{Vinaya K.\ Kavatamane}
\affiliation{Institute for Quantum Science and Technology, University of Calgary, Calgary, AB, T2N 1N4, Canada}
\author{Matthew Mitchell}
\altaffiliation{Currently with Dream Photonics Inc., Vancouver, BC, V6T 1Z4, Canada}
\affiliation{Institute for Quantum Science and Technology, University of Calgary, Calgary, AB, T2N 1N4, Canada}
\author{David P.\ Lake}
\altaffiliation{Currently with the Applied Physics Department, California Institute of Technology, Pasadena, CA 91125, USA}
\affiliation{Institute for Quantum Science and Technology, University of Calgary, Calgary, AB, T2N 1N4, Canada}
\author{Paul E.\ Barclay}
\affiliation{Institute for Quantum Science and Technology, University of Calgary, Calgary, AB, T2N 1N4, Canada}

\date{\today}
        
\begin{abstract}
Diamond's unique physical properties make it a versatile material for a wide range of nonlinear and quantum photonic technologies. 
However, unlocking diamond's full potential as a nonlinear photonic material with non-zero second-order susceptibility $\chi^{(2)}\neq0$ requires symmetry breaking. In this work, we use a nanoscale cavity to demonstrate second-harmonic generation (SHG) in diamond, and demonstrate, for the first time, that the magnitude of the diamond's effective $\chi^{(2)}$ strongly depends on the electronic configuration of defects in the diamond crystal, such as nitrogen-vacancy centres. 
The modification of $\chi^{(2)}$ arises from photoionisation from the negative to neutral charge-state, and is manifested by quenching of SHG upon green illumination.
Toggling the green illumination allows for optical switching of the device's $\chi^{(2)}$. Optical control of $\chi^{(2)}$ by defect engineering opens the door for second-order nonlinear processes in diamond. 
\end{abstract}
                      
\maketitle
Diamond's exceptional physical properties make it an attractive platform for many photonics applications\,\cite{Shandilya2022}. Diamond photonic devices play a key role in interfacing light with colour centres used in quantum technologies\,\cite{Janitz2020,Hedrich2020,Pasini2024}, as well as nanomechanical sensors and information processing devices\,\cite{Shandilya2021, Joe2024}. However, diamond's centrosymmetric crystal structure and vanishing second-order nonlinearity (${\chi^{(2)}=0}$) limit its use in nonlinear optics to applications involving third-order processes\,\cite{Hausmann2014,Riedel2020,Flagan2022Dres}. Spatial symmetry breaking enables otherwise forbidden second-order interactions in centrosymmetric materials\,\cite{Cazzanelli2016}, as studied in 2D\ materials\,\cite{Wang2015PRL, Huang2024AdvFunMat}, nanomaterials\,\cite{Liu2018NatPhys, Zhao2022AdvPhotRes}, and surface science\,\cite{Buck1995, Lupke1999, Shi2016PhysRevB}, and underlie complex nonlinear optical effects in nanophotonic devices\,\cite{Galli2010, Chen2024NanoLett}.
Prominent examples include photoinduced second-harmonic generation (SHG) mediated by the coherent photogalvanic effect\,\cite{Anderson1991, Dianov1995, Nitiss2020, Yakar2022} in Si$_3$N$_4$\,\cite{Porcel2017, Billat2017, Hickstein2019,Lu2021, Lu2021PRAppl, Nitiss2022, Clementi2025}. 

Defects in centrosymmetric crystals are a well-known mechanism for inducing $\chi^{(2)}\neq0$\,\cite{Li2022CellReports}, and can be used to probe the defect density\,\cite{Lim2000, Fiore2011, Wang2017NatCom}. Furthermore, static electric fields ($E_{\textrm{DC}}$), applied externally\,\cite{Terhune1962,Lee1967,Timurdogan2017,Widhalm2022} or generated internally by charged defects\,\cite{Schriever2015,Castellan2019,Zhang2023ACSPhot}, can couple to the third-order susceptibility ($\chi^{(3)}$) resulting in electric-field-induced second-harmonic generation (EFISH). 
In this third-order process\,\cite{Cai2011}---previously demonstrated in silicon\,\cite{Schriever2015,Castellan2019} and double
perovskites\,\cite{Zhang2023ACSPhot}---the induced second-order nonlinearity is\,\cite{Schriever2015,Castellan2019}
\begin{equation}
\chi_{\textrm{EFISH}}^{(2)}=3\chi^{(3)}E_{\textrm{DC}}\,. \label{eq:chi_EFSH}
\end{equation}

In centrosymmetric materials, the effective second-order susceptibility, $\chi_{\textrm{eff}}^{(2)}$, is the sum of contributions\,\cite{Motojima2019,Li2022CellReports}
\begin{equation}
    \chi_{\textrm{eff}}^{(2)} =  \chi_{\textrm{host}}^{(2)} + \sum\chi^{(2)}_{\textrm{defects}}\,,
\end{equation}
where $\chi^{(2)}_{\textrm{defects}}$ is induced by crystal defects and $\chi^{(2)}_{\textrm{host}}$ is from the host device, encompassing surface effects\,\cite{Levy2011,Zhang2019} and electric field variations responsible for EFISH\,\cite{Puckett2016, Timurdogan2017}.

Given the prominence of defects in diamond quantum photonics---for example nitrogen-vacancy (NV) centres---understanding how defects induce second-order nonlinearities in diamond is imperative. Recently, SHG has been observed in diamond using ultrafast laser pulses interacting with shallow (35 nm deep) high density ($3\times10^{17}\,\textrm{cm}^{-3}$)  NV centres layers\,\cite{Abulikemu2021, Abulikemu2022}, suggesting that they enhance $\chi^{(2)}_\text{eff}$.
Here, we probe this effect by demonstrating, for the first time, that $\chi^{(2)}_\text{eff}$ changes when the charge-state of NV centres is optically modified, directly confirming that NVs can enhance $\chi^{(2)}_\text{eff}$.
Using a diamond nanophotonic cavity that generates SHG using a low-power (mW) continuous wave (CW) infrared (IR) laser, we show that $\chi^{(2)}_\text{eff}$ is modified by excitation of NV centres with an independent laser field. By comparing the SHG modification with the NV photoluminescence (PL) from the optical excitation, and by building on advances in understanding NV charge-state dynamics obtained using similar cavities \cite{Shandilya2024arxiv}, we firmly establish a correlation between the NV charge-state and $\chi^{(2)}_\text{eff}$.
This effect allows all-optical control of diamond's optical nonlinearity, and provides insight into the mechanisms connecting NV centres and $\chi^{(2)}$. Furthermore, we show that this effect can probe optical charge-state conversion of NV centres.

\begin{figure}[t]
\centering
\includegraphics[width=\columnwidth]{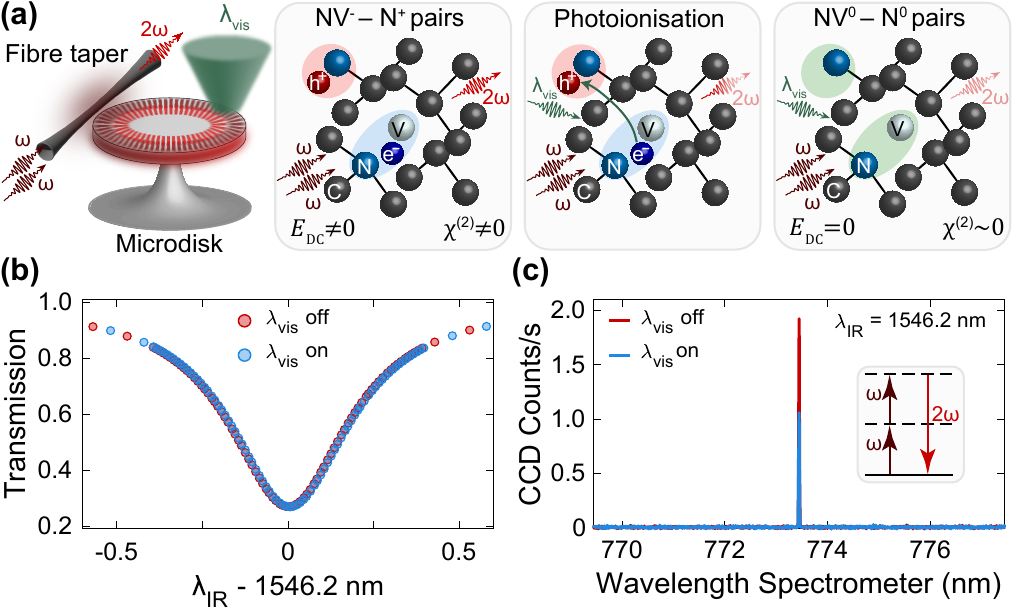}
\caption{\textbf{(a)} (\textit{left}) Schematic of the fibre-taper-coupled diamond microdisk used to demonstrate optically modulated SHG. The spot-size of the green laser field is not to scale. (\textit{right}) A proposed mechanism for the observed behaviour of $\chi^{(2)}$. An electric field ($E_\text{DC}$) created by charged crystal-defects induces $\chi_{\text{eff}}^{(2)}\neq0$. Photoionisation by a green laser ($\lambda_\text{vis}$) combined with the IR field modifies $\chi^{(2)}_{\text{eff}}$, leading to quenching of the SHG.
\textbf{(b)} Normalised fibre-taper transmission spectrum of the IR mode, showing that the cavity is unaffected by the green laser.
\textbf{(c)} In burgundy, SHG spectrum when the IR laser is resonant with the cavity mode in (b). In blue, the SHG signal is quenched when the green laser is on. The inset shows the mechanism of SHG: two photons at frequency $\omega$ upconvert to a photon at frequency $2\omega$.}
\label{Fig:fig_1}
\end{figure}

The device studied here is a microdisk fabricated from optical-grade diamond that hosts uniform distributions of NV centres (density $\sim10^{13}-10^{14}\,\textrm{cm}^{-3}$\,\cite{Acosta2009}) and substitutional nitrogen (N$_{\text{s}}$, density $\sim10^{17}\,\textrm{cm}^{-3}$). As illustrated in Fig.\,\ref{Fig:fig_1}\,(a), the microdisk supports whispering-gallery modes with quality factors $Q = 10^3-10^5$\,\cite{Masuda2024} evanescently coupled to a fibre-taper waveguide. Second-harmonic generation--upconversion of two photons at frequency $\omega$ to one photon at $2\omega$--is studied using a tunable CW laser ($P_{\textrm{IR}}=76\,\textrm{mW}$) coupled to a  microdisk mode (wavelength $\sim1546\,\textrm{nm}$) via the fibre-taper\,\cite{Lake2016}.
To enable optical control of defects in the microdisk, it is positioned in the focus of a diffraction limited confocal microscope ($\text{NA}=0.8$) that excites NV centre PL with a green ($\lambda_\text{vis} = 532\,\text{nm}$) CW laser\,\cite{Masuda2024}.
This system can be used to study the effect of IR fields on the microscopic properties of NV centres, as shown recently in Ref.\,\cite{Shandilya2024arxiv}, and to probe the effect of NVs on the optical properties of diamond, as described in this Letter.

\begin{figure}[t]
\centering
\includegraphics[width=\columnwidth]{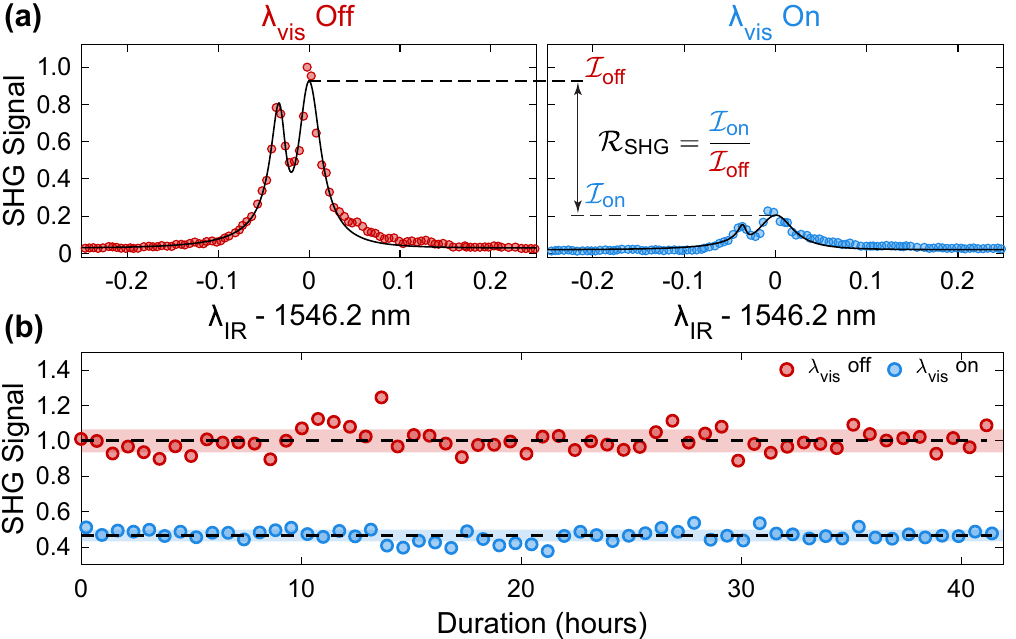}
\caption{Deterministic optical switching of $\chi^{(2)}_{\text{eff}}$ by illuminating the microdisk with a green laser.
\textbf{(a)} Dependence of SHG on IR laser detuning from cavity resonance in the absence (burgundy) and presence (blue) of the green laser. A quenching of $\sim80\,\%$ is observed from the green laser. Data in blue has been corrected for background PL from NV centres. 
\textbf{(b)} Toggling of $\chi^{(2)}_{\text{eff}}$ by performing successive SHG measurement with (blue) and without (burgundy) green excitation. The quenching remains constant over a duration of 40 hours. Both datasets are normalised to the mean value  with the green laser off. The dashed black lines and the shaded regions correspond to the mean and the standard deviation for the respective dataset.}
\label{Fig:fig_2}
\end{figure}

\begin{figure}[t!]
\centering
\includegraphics[width=\columnwidth]{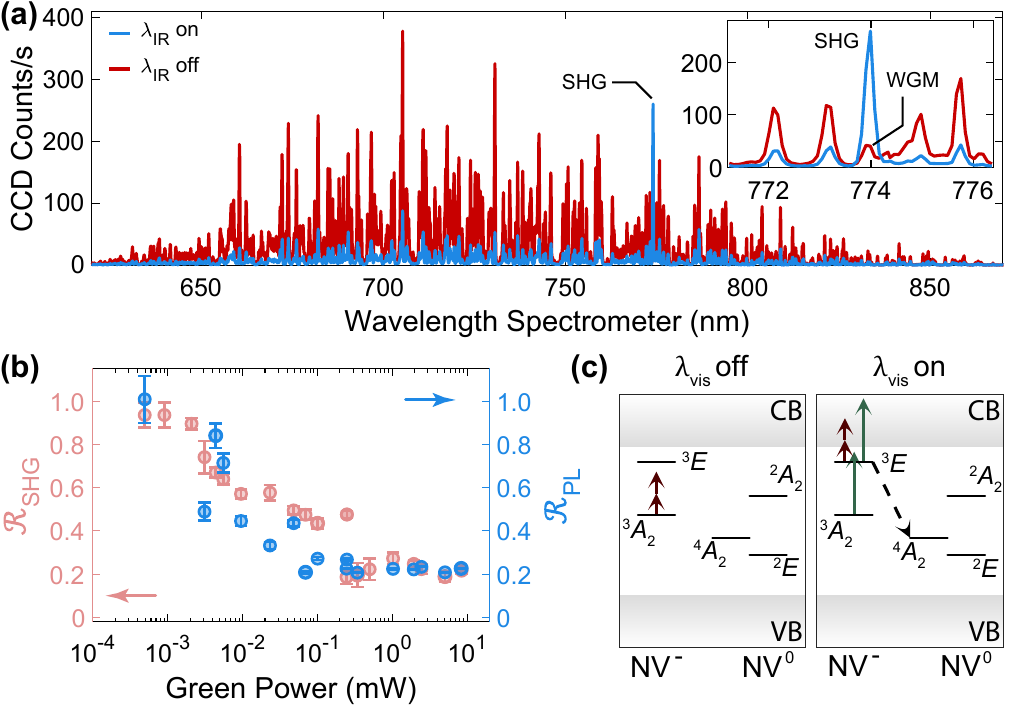}
\caption{Charge-state conversion.
\textbf{(a)} Fibre-taper collected NV centre PL spectrum from green excitation of the microdisk. Coupling an IR field into the microdisk suppresses the PL (in blue).
The inset showcases that the SHG signal is resonant with a whispering-gallery mode (WGM).
\textbf{(b)} SHG quenching (pink, left axis) and suppression of NV centre PL (blue, right axis) for varying green laser power. A monotonic decrease is observed for both PL suppression and SHG quenching, suggesting a similar underlying mechanism.
\textbf{(c)} Position of the relevant energy levels of NV$^0$ and NV$^-$ drawn to scale. The burgundy and green arrows represent IR and green photons, respectively. 
\textit{(left)} Under IR excitation, higher-order photon ($>2$) processes are required to excite NV$^-$ and population remains in NV$^-$. 
\textit{(right)} Green photons populate the $^3E$ excited state of NV$^-$, from where two IR photons pump the population into the dark $^4A_2$ state of NV$^0$ (black dashed arrow).
}
\label{Fig:fig_3}
\end{figure}

The fibre-taper transmission spectrum of the IR laser scanned across a mode is shown in Fig.\,\ref{Fig:fig_1}\,(b), and a SHG spectrum collected by the fibre-taper when the IR laser wavelength $\lambda_\text{IR}$ is on-resonance is shown in Fig.\,\ref{Fig:fig_1}\,(c).
To observe the otherwise weak SHG signal, we leverage the broad mode spectrum of the microdisk, which supports resonances across visible and IR wavelengths\,\cite{Masuda2024}.
The SHG process is enhanced by modes at $\lambda_\text{IR}$ and $\lambda_{\textrm{IR}}/{2}$, as shown in Fig.\,\ref{Fig:fig_2}\,(a), where the SHG intensity is recorded as $\lambda_\text{IR}$ is tuned through the resonance (Fig.\,\ref{Fig:fig_1}\,(b)). Second-harmonic generation is only observed with the IR field near resonance, is strongest where $\lambda_{\textrm{IR}}/{2}$ is resonant with a mode, and exhibits a double-peak structure due to nearly-degenerate modes near the SHG wavelength. 
Further evidence of the doubly-resonant enhancement is evident from sharp features in the NV PL spectrum at mode wavelengths\,\cite{Masuda2024}, including a peak at the SHG wavelength (Fig.\,\ref{Fig:fig_3}\,(a)). Its relatively small amplitude may result from poor coupling between the highest-$Q$ microdisk modes and the fibre-taper\,\cite{Masuda2024}.

To probe whether the SHG is related to NV centres, we excite NVs in the microdisk with the green laser while monitoring the SHG. We observe strong SHG quenching by the green laser, as shown in Figs.\,\ref{Fig:fig_1}\,(c) and \ref{Fig:fig_2}\,(a). In these measurements, the green laser is focused on the microdisk edge where whispering-gallery modes are localised (Fig.\,\ref{Fig:fig_1}\,(a)).
We note that for large green laser power, the SHG quenching  was insensitive to spot-size and position, provided it spatially overlaps with the whispering-gallery modes.
To rule out modification of SHG by laser-induced shifting of mode wavelengths, we monitor the IR pump mode (Fig.\,\ref{Fig:fig_1}\,(b)) and a cavity-enhanced third-harmonic signal (not shown) with and without green laser excitation, and observe no change to either, confirming that the green laser's effect on $\chi^{(2)}_{\text{eff}}$ is responsible for the SHG quenching. We repeated this measurement over $40\,\text{hours}$ while toggling the green laser (Fig.\,\ref{Fig:fig_2}\,(b)). In the absence of green illumination, the SHG signal recovers, demonstrating deterministic switching of the device's $\chi^{(2)}_{\text{eff}}$, a previously unreported effect, to the best of our knowledge.

The SHG quenching points to defects such as NV centres being key contributors to  $\chi^{(2)}_{\text{eff}}$.
Two optical processes modifying the NV centre electronic state can affect $\chi^{(2)}_{\textrm{NV}}$, the second-order nonlinearity of the NV centres.
First, green excitation cycles the predominantly negatively charged NV$^-$ between $^3A_2$ ground- and $^3E$ excited states\,\cite{Gali2008}, generating the PL spectra in Fig.\,\ref{Fig:fig_3}\,(a). The average change in electron configuration during this cycling may change $\chi^{(2)}_{\textrm{NV}}$; the SHG quenching would imply a smaller $\chi^{(2)}_{\textrm{NV}}$ for $^3E$ than $^3A_2$.
Second, the combination of green and IR fields can photoionise NV$^-$ to a NV$^0$ dark state\,\cite{Shandilya2024arxiv}. This process is illustrated  in Fig.\,\ref{Fig:fig_3}\,(c) and revealed experimentally in Fig.\,\ref{Fig:fig_3}\,(a) by the IR field's suppression of NV PL.
This two-step process requires excitation to  $^3E$, followed by photoionisation to the $^4A_2$ state of NV$^0$ via a single green or two IR photons.
For the large IR intensity in the microdisk, the latter two-photon photoionisation mechanisms dominates and traps population in the dark $^4A_2$ state\,\cite{Shandilya2024arxiv}. This explanation implies that NV$^0$ possesses a smaller $\chi^{(2)}_{\textrm{NV}}$ than NV$^{-}$, or that the change to the local charge environment when NV$^-$ is converted to NV$^0$ affects $\chi_{\textrm{eff}}^{(2)}$.

To test these explanations, we measure the effect of green excitation power on SHG quenching (Fig.\,\ref{Fig:fig_3}\,(b)).
We define the relative SHG intensity as $\mathcal{R}_{\textrm{SHG}}={\mathcal{I}_{\textrm{on}}}/{\mathcal{I}_{\textrm{off }}}$, where $\mathcal{I}_{\textrm{on\,(off)}}$ is the SHG peak intensity with (without) green excitation. We also plot the corresponding suppression of NV PL, defined as $\mathcal{R}_{\textrm{PL}}={\mathcal{L}_{\textrm{on}}}/{\mathcal{L}_{\textrm{off}}}$ where $\mathcal{L}_{\textrm{on\,(off)}}$ is PL integrated over the NV$^-$ emission band ($640$--$770$\,nm) in the presence (absence) of the IR field. The strong correlation between $\mathcal{R}_{\textrm{SHG}}$ and $\mathcal{R}_{\textrm{PL}}$ suggests that the SHG strength is related to the degree of IR photon-assisted photoionisation to NV$^0$ observed in Ref.\ \cite{Shandilya2024arxiv}. This provides conclusive evidence that the electronic configuration of crystal defects plays a critical role in determining their influence on the optical properties of the host crystal. Note that when measuring $\mathcal{R}_{\textrm{SHG}}$ we account for PL emitted into the SHG mode by first measuring PL in absence of the IR field, and then assuming that it is suppressed by the IR field in equal proportion as PL into other nearby modes (inset Fig.\,\ref{Fig:fig_3}\,(a)).
In these measurements, the green laser power, measured immediately before the objective, is scattered by the microdisk into a complex intensity distribution. Green light output from  the fibre-taper suggests  that some is scattering into whispering-gallery modes.

As introduced above, the observed dependence of $\chi^{(2)}_{\text{eff}}$ on NV charge-state can arise from several mechanisms. Most directly, the microscopic $\chi^{(2)}_{\textrm{NV}}$ of the NV$^0$ electronic structure may be smaller than that of NV$^-$; this is consistent with \textit{ab initio} calculations and is related to the larger spacing between the ground and excited states of NV$^0$ than NV$^-$\,\cite{Li2022CellReports}. In addition, the local charge environment and electric field created by the NV$^-$ ensemble could induce a $\chi^{(2)}_\text{host}$ that is quenched when NV$^-$ is converted to NV$^0$. For example, NV$^-$+N$_{\text{s}}^+\rightarrow$NV$^0$+N$_{\text{s}}^0$ reduces the local charge environment by two\,\cite{Manson2018,Blakley2024}---this reduction in $E_{\text{DC}}$ diminishes $\chi^{(2)}_{\text{eff}}$ according to Eq.\,\ref{eq:chi_EFSH}. Probing such changes in solid-state environments via SHG intensity is an application of this effect, which, as illustrated in Fig.\,\ref{Fig:fig_1}\,(a), may be affected by the density of other impurities, such as substitutional nitrogen\,\cite{Ashfold2020,Goldblatt2024arxiv}, and the device geometry and its effect on electron accumulation. Further measurements, particularly in the time domain, are required to definitively identify the dominant mechanism.

\begin{figure}[bt!]
\centering
\includegraphics[width=\columnwidth]{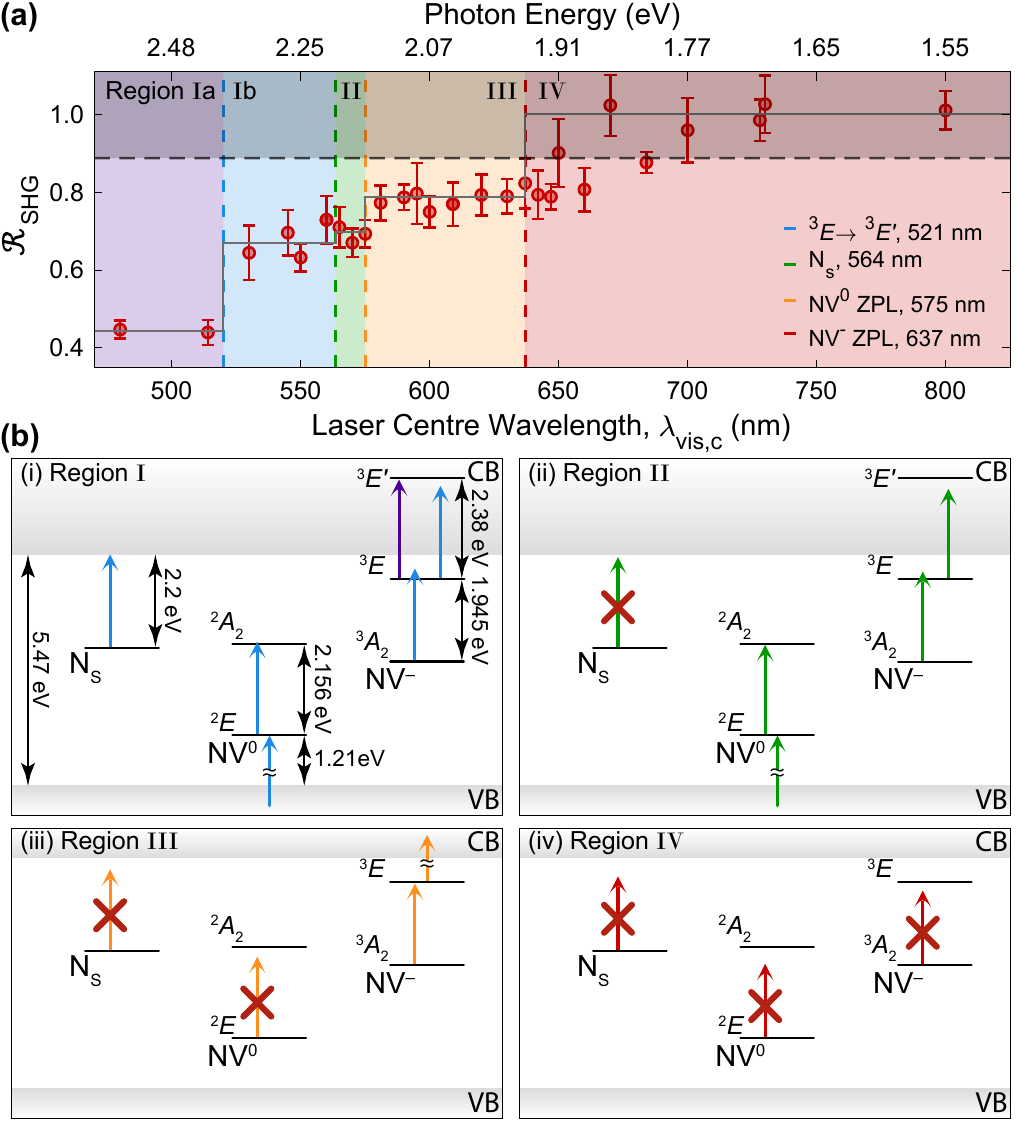}
\caption{Dependence of $\mathcal{R}_{\text{SHG}}$ on visible excitation wavelength, $\lambda_{\text{vis,c}}$. \textbf{(a)} A monotonic decrease in $\mathcal{R}_{\text{SHG}}$ is observed for shorter $\lambda_{\text{vis,c}}$.
The blue dashed line separating Region\,\Romannum{1}a and \Romannum{1}b marks the energy requirement for photoionisation from $^3E$ to the $^3E'$ state located $2.38\,\text{eV}$ higher.
Dashed green, orange and red lines represent the photoionisation threshold for N$_{\text{s}}$ and the zero-phonon lines for NV$^{0}$ and NV$^{-}$, respectively.
Photoionisation of N$_{\text{s}}$ and both NV charge-states is energetically allowed in Region\,\Romannum{1}, resulting in large SHG quenching (small $\mathcal{R}_{\text{SHG}}$). The plateaux in $\mathcal{R}_{\text{SHG}}$ observed in Region\,\Romannum{2} and \Romannum{3}, signifies regimes where optical excitation of N$_{\text{s}}$ and NV$^0$, respectively, is energetically forbidden. In Region\,\Romannum{3}, only photoionisation of NV$^-$ is energetically allowed.
For $\lambda_{\text{vis,c}}>637\,\textrm{nm}$ (Region\,\Romannum{4}), all photoionisation processes are prohibited, and no quenching is observed.
The grey shaded region indicates the experimental uncertainty set by the standard deviation in measurements without visible excitation. Horizontal grey lines are guides to the eye. 
\textbf{(b)} The relevant energy levels for N$_\text{s}$, NV$^0$ and NV$^-$ within diamond's bandgap. The sub-panels indicate permitted photoionisation processes for each region in (a). 
For simplicity, the state $^3E'$ is omitted in panel (iii) and (iv) as the energy required to populate this state far exceeds the available photon energies in Regions\,\Romannum{3} and \Romannum{4}.} 
\label{Fig:fig_4}
\end{figure}

To elucidate the contribution from charged defects to $\chi^{(2)}_{\text{host}}$, we measured the dependence of $\mathcal{R}_{\textrm{SHG}}$ on $\lambda_\text{vis}$ by substituting the green laser with a pulsed supercontinuum laser (centre wavelength $\lambda_{\text{vis,c}}$) and measuring $\mathcal{R}_{\textrm{SHG}}$ for $\lambda_{\text{vis,c}}=480\dots800\,\text{nm}$ ($2.58\dots1.55\,\text{eV}$). 
The SHG quenching depends strongly on $\lambda_{\text{vis,c}}$. It is large for $\lambda_{\text{vis,c}}<564\,\text{nm}$ (Region\,\Romannum{1} in Fig.\,\ref{Fig:fig_4}\,(a)), and no quenching is observed for $\lambda_{\text{vis,c}}\gtrsim650\,\text{nm}$ (Region\,\Romannum{4}).
Each data point in Fig.\,\ref{Fig:fig_4}\,(a) is the average peak SHG value from three successive IR laser scans across resonance, with visible excitation on, normalised to the same measurement with visible excitation off.
Error bars are the standard deviation from the repeated measurements followed by standard error propagation methods. The shaded range for $\mathcal{R}_{\textrm{SHG}}> 0.87$ indicates the experimental uncertainty from the standard deviation and error propagation obtained from repeated measurements with the visible excitation off; SHG quenching cannot be reliably resolved within this region.  

To explain the dependence of $\mathcal{R}_{\textrm{SHG}}$ on $\lambda_{\text{vis,c}}$, we consider the energy levels of N$_{\text{s}}$ and the NV centre within diamond's bandgap in Fig.\,\ref{Fig:fig_4}\,(b). For $\lambda_{\text{vis,c}}<564\,\text{nm}$, where the largest quenching is observed (Region\,\Romannum{1}), the visible excitation laser photon energy photoionises N$_{\text{s}}$ (threshold $2.2\,\text{eV}$\,\cite{Rosa1999,Orphal-Kobin2023}) and both NV charge-states (Fig.\,\ref{Fig:fig_4}\,(b,\romannum{1})), resulting in large modifications of the charge environment\,\cite{Siyushev2013,Wolters2013}.
Electrons liberated from photoionisation of N$_{\text{s}}$ can be captured by NV$^0$, constituting an additional recombination mechanism from NV$^0$ to NV$^-$, i.e.\ N$_{\text{s}}^0+$NV$^0\rightarrow$N$_{\text{s}}^{+}+$NV$^-$\,\cite{Todenhagen2025}, and creating larger NV$^-$ and N$_{\text{s}}^+$ populations than expected solely from optically-driven charge-state cycling between NV$^0$ and NV$^-$\,\cite{Todenhagen2025}. The behaviour of $\mathcal{R}_{\textrm{SHG}}$ within Region\,\Romannum{1}b can therefore be explained by residual electric fields generated by these NV$^-$--N$_{\text{s}}^{+}$ pairs\,\cite{Manson2018}.

The abrupt reduction in $\mathcal{R}_{\textrm{SHG}}$ for $\lambda_{\text{vis,c}}=520\,\textrm{nm}$ (Region\,\Romannum{1}a) can be explained by photoionisation from the $^3E$ excited state of NV$^{-}$ to a recently confirmed spin-triplet excited state, $^3E'$, located $2.38\,\textrm{eV}$ higher in the conduction band\,\cite{Luu2025}. Although little is known about the $^3E'$ state, signatures of its existence have been observed in absorption\,\cite{Kupriyanov2000}, photoluminescence excitation\,\cite{Beha2012,Todenhagen2025} and transient absorption spectroscopy\,\cite{Luu2024,Luu2025}. From $^3E'$ the electron is rapidly lost into the conduction band\,\cite{Beha2012}, increasing the NV$^{0}$ population. The strong quenching of SHG at $\lambda_{\text{vis,c}}=520\,\textrm{nm}$ is further testimony to the presence of the $^3E'$ state.

The available photon energies in this work restricts photoionisation from NV$^-$ to NV$^0$ to occur from the $^3E$ state. Promoting population from $^3A_2$ to $^3E$ requires photon energies $>1.945\,\text{eV}$ ($637\,\text{nm}$). Consequently, for $\lambda_{\text{vis,c}}>637\,\text{nm}$ (Fig.\,\ref{Fig:fig_4}\,(b,\romannum{4})), photoionisation is energetically forbidden and the charge environment is unchanged, explaining the absence of quenching in Region\,\Romannum{4}.

Recombination from NV$^0$ to NV$^-$ occurs from the $^2A_2$ state of NV$^0$. Promoting population from $^2E$ to $^2A_2$ requires photon energies $>2.156\,\text{eV}$ ($575\,\text{nm}$)\,\cite{Manson2013}. For $564\,\text{nm}<\lambda_{\text{vis,c}}<575\,\text{nm}$ (Region\,\Romannum{2}), photoionisation and recombination between the two charge-states is possible (Fig.\,\ref{Fig:fig_4}\,(b,\romannum{2})).
Crucially, in this wavelength range, these processes occur without optical excitation of N$_{\text{s}}$, resulting in substantially smaller changes to the charge environment\,\cite{Siyushev2013, Chu2014, Pfaff2014, Orphal-Kobin2023}, and larger $\mathcal{R}_{\textrm{SHG}}$ in Region\,\Romannum{2} and \Romannum{3} compared to that of Region\,\Romannum{1}.

Finally, for $575\,\text{nm}<\lambda_{\text{vis,c}}<637\,\text{nm}$ (Region\,\Romannum{3}), only excitation and photoionisation of NV$^-$ is energetically permitted (Fig.\,\ref{Fig:fig_4}\,(b,\romannum{3})). While direct single-photon recombination from NV$^0$ is prohibited\,\cite{Beha2012}, weak anti-Stokes assisted excitation of $^2E\rightarrow{^2A_2}$ may contribute to charge-state cycling\,\cite{Wood2024}. Trapping the system in NV$^0$ explains the constant quenching observed in Region\,\Romannum{3}. Note that  differences in the quenching between the green CW experiment in Fig.\,\ref{Fig:fig_2} and Fig.\,\ref{Fig:fig_4} alongside deviations from the idealised energy transitions discussed here can be attributed to the pulsed nature and the broad bandwidth of the supercontinuum laser: the IR laser is always present, whereas the supercontinuum laser has a pulse duration of $6\,\text{ps}$, a repetition rate of $80\,\text{MHz}$  and a spectral bandwidth of $\sim10\,\text{nm}$. 

The above discussion is centred around N$_{\text{s}}^{+}$ and NV$^{-}$ being the dominant contributors to $\chi^{(2)}_{\text{eff}}$. 
The motivation behind this is twofold. First, nitrogen is the most prevalent impurity in diamond\,\cite{Ashfold2020,Todenhagen2025}. Second, we observe a strong correlation between $\mathcal{R}_{\textrm{PL}}$ and $\mathcal{R}_{\textrm{SHG}}$ (Fig.\,\ref{Fig:fig_3}\,(b)) alongside excellent concordance between the step-wise change of $\mathcal{R}_{\textrm{SHG}}$ with $\lambda_{\text{vis,c}}$ at the photoionisation thresholds for N$_{\text{s}}$ and the NV centres\,\cite{Orphal-Kobin2023}. 
Nevertheless, we cannot exclude the contribution from other crystal defects, such as mono- and divacancies\,\cite{Clark1973,Pu2001,Gorlitz2022}, hydrogen related defects\,\cite{Rosa1999,Czelej2018,Day2024} and additional nitrogen-vacancy complexes\,\cite{Ashfold2020,Deak2014}.

Our work is the first demonstration of second-order nonlinear integrated photonics in diamond. However, a thorough characterisation of $\chi^{(2)}$ is needed to fully harness diamond's potential for nonlinear optics. Quantifying the magnitude of $\chi^{(2)}$ from SHG requires measuring the internal conversion efficiency, which we are unable to perform without a tunable laser around $774\,\text{nm}$ to characterise the SHG to fibre-taper coupling efficiency.
In future, the magnitude of $\chi^{(2)}_{\text{eff}}$ can be deduced from Maker fringes\,\cite{Maker1962,Jerphagnon1970}, or two-photon imaging\,\cite{Nitiss2022}. However, the latter approach requires calibration to a known $\chi^{(2)}$, e.g.\ from the conversion efficiency of electrically poled waveguides\,\cite{Timurdogan2017}. 
Finally, measuring a cavity resonance shift induced by externally applied electric fields could provide insight into $\chi^{(2)}_{\text{eff}}$ through its relationship to the DC Kerr effect\,\cite{Chang2022APL,Shetewy2025}.

Crystal defects are ubiquitous in wide-bandgap semiconductors.
Our findings are therefore relevant to enabling $\chi^{(2)}\neq0$ processes in materials via defect engineering. For example, liberation and diffusion of charges is essential to SHG mediated by the coherent photogalvanic effect\,\cite{Anderson1991,Yakar2022,Porcel2017, Nitiss2020}.  

In conclusion, we have, for the first time, demonstrated SHG from a diamond nanophotonic cavity.
We directly show that the device's $\chi^{(2)}_{\text{eff}}$ strongly depends on the charge-state of defects native to the diamond. By selectively exciting the microdisk with a visible laser, we demonstrate, for the first time in a nanophotonic device, deterministic modulation of the magnitude of $\chi^{(2)}_{\text{eff}}$, which we attribute to changes in the local electric field related to the NV charge state. In other words, the electric configuration of crystal defects directly modifies the optical properties of the host diamond. Monitoring the strength of the SHG signal is also a novel approach to sense changes in the solid-state environment of the cavity. A deeper understanding of the relationship between $\chi^{(2)}_{\text{eff}}$ and the charge-state provides the foundation for frequency conversion, modulators, and optical switching via $\chi^{(2)}$ engineering, opening the door for second-order nonlinear processes--a valuable addition to the diamond photonics toolbox.

\begin{acknowledgments}
This work was supported by NSERC (Discovery Grant and Alliance Quantum programs), Alberta Innovates, and the Canadian Foundation for Innovation. SF acknowledges support from the Swiss National Science Foundation (Project No. P500PT\_206919).
\end{acknowledgments}

\appendix

%\section{Appendixes}

%\bibliography{Optical_Switching_NVc_Revision_1}% Produces the bibliography via BibTeX.


\end{document}